\documentclass[a4paper]{article}
\usepackage{INTERSPEECH2020}
\usepackage{multirow}
\usepackage{bbding}

\title{Multi-Label Training for Text-Independent Speaker Identification}
\name{Yuqi Xue}
\address{Department of Acoustical Science and Engineering, University of Nanjing, Nanjing, China}
\email{yuqixue@smail.nju.edu.cn}

\begin{document}

\maketitle
\begin{abstract}
  In this paper, we propose a novel strategy for text-independent speaker identification system: Multi-Label Training (MLT). Instead of the commonly used one-to-one correspondence between the speech and the speaker label, we divide all the speeches of each speaker into several subgroups, with each subgroup assigned a different set of labels. During the identification process, a specific speaker is identified as long as the predicted label is the same as one of his/her corresponding labels. We found that this method can force the model to distinguish the data more accurately, and somehow takes advantages of ensemble learning, while avoiding the significant increase of computation and storage burden. In the experiments, we found that not only in clean conditions, but also in noisy conditions with speech enhancement, Multi-Label Training can still achieve better identification performance than commom methods. It should be noted that the proposed strategy can be easily applied to almost all current text-independent speaker identification models to achieve further improvements.

\end{abstract}
\begin{figure*}[ht]
	\centering
	\includegraphics[width=110mm]{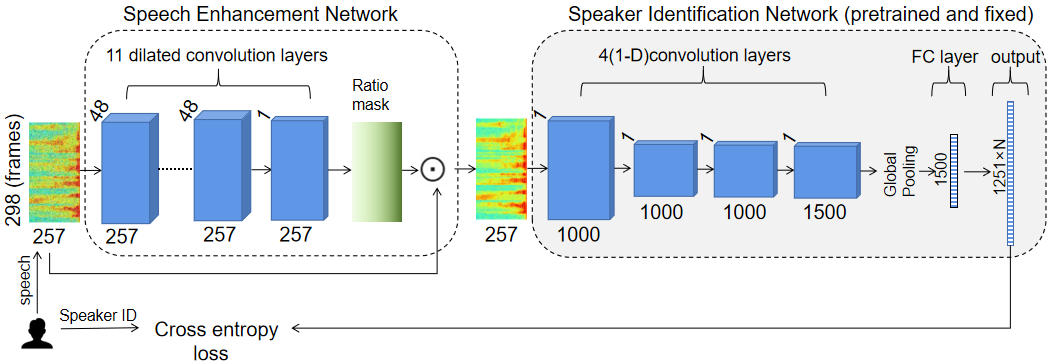}
	\caption{A flow chart for the Multi-Label Training(N means the number of subgroups)}
	\label{chart}
\end{figure*}
\section{Introduction}
Recently, deep learning has greatly advanced the development of various speech technologies \cite{graves2006connectionist, deng2021improving, wang2021fine, DBLP:conf/interspeech/DengCM21, DBLP:conf/interspeech/Deng0SA22, poddar2018speaker, deng2024lst, chen2022large}. Among which speaker identification is a task that recognizes the identity of the speaker from their speech characteristics \cite{poddar2018speaker, vaessen2022fine, cai2023pretraining}. In recent years, the use of deep learning techniques \cite{lecun2015deep} has significantly improved the performance of speaker identification. Variani et al. proposed the use of multiple layers of fully connected neural networks to form d-vectors \cite{variani2014deep}, and Snyder et al. proposed X-vectors based on time-delayed neural networks \cite{snyder2018x}. Using a large number of enhanced data sets, the speaker embedding approaches \cite{snyder2017deep, heigold2016end} can now surpass the traditional i-vector approach \cite{dehak2010front}. A number of other methods for extracting robust embedding from speech have been proposed after that. Voxceleb \cite{nagrani2017voxceleb,chung2018voxceleb2}, a free and open source dataset, provides a bridge for comparison between different methods, thus facilitating the advances in speaker recognition. However, speaker recognition in noisy conditions is still a challenging task.

There are now many solutions for noisy speech. Some studies \cite{leglaive2019speech, sadeghi2020audio} have attempted to recover the original speech by speech enhancement, others \cite{jang2017enhanced, DBLP:conf/mrcs/FarahaniAH06} have tried to extract features from clean speech, and some works \cite{DBLP:journals/ejasmp/NahmaYDN19, DBLP:journals/ejasp/YaoZZ16} have attempted to estimate speech quality by calculating SNR. For voice enhancement, most of the previous works were done independently. But enhanced speech is not always suitable for speech recognition \cite{novoselov2022robust}, while speech suitable for speech recognition does not always sound clean, so it is necessary to train the enhancement and recognition model jointly. E2E methods that jointly learn acoustic and linguistic knowledge \cite{10572334, DBLP:conf/interspeech/ShonTG19, DENG2024103109} have achieved great success recently.
Shon et al. attempted to integrate the speech enhancement model and the speaker recognition model using one framework \cite{DBLP:conf/interspeech/ShonTG19} in which the enhancement model filters out unwanted features by ratio masking and multiplying it point-by-point with the input spectrum. They adopt their new loss function, which differs from the traditional L2 paradigm loss of enhanced spectrum and reference spectrum.

DNN is also an important topic when applied to speaker identification. Since the performance of DNN for speaker identification is not the deeper the better, Rohdin et al. \cite{DBLP:journals/csl/RohdinSDPMBG20} proved that the model did not gain advance from its more complex structure after adopting a deeper and more complex deep neural network. The number of hidden layers and nodes has a strong influence on DNN, and the optimal number of hidden layers and nodes is closely related to the dimension of the input and output layer, but there is no consensus on how to determine them \cite{Rachmatullah2019DeterminingTN, 10447838}. Nagrani et al. \cite{nagrani2017voxceleb} trained the residual networks on the Voxceleb1 and Voxceleb2 respectively. Since these two data sets contain different numbers of speakers and therefore need different output layer dimensions, they trained the models in which the last hidden layer's the dimension was nearly proportional to the output layer's dimension on these two data sets.

Ensemble learning is an important area of machine learning, which is used in many different ways and constantly refined. Considering the complexity of the ensemble system and the recognition rate of each classifier are always opposed, Mao et al. \cite{DBLP:journals/asc/MaoCJGW19} proposed a transformed ensemble learning system. After redefining the complexity, they used a new function that takes into account both the system complexity and the individual classifiers' recognition rate. When training, a random batch of data from the original training set was selected as a subset, and then the corresponding classifier was trained on it. Agarwal et al. \cite{DBLP:journals/eswa/AgarwalC20} consider that the performance of ensemble learning greatly depends on the complexity between individual classifiers, while the training subsets used in some traditional ensemble learning methods are not completely mutually exclusive and the correlation is high, thus resulting in a low system complexity. So they proposed the A-Stacking and A-Bagging ensemble method, which first clustered the original training set and then split it into numerous subsets that are mutually exclusive to each other, improving the complexity of the system, and then trained the individual classifiers on the corresponding subsets.

While superior model structures can increase speaker identification results, the superiority of ensemble learning is clear for all to see, and this paper attempts to draw on some advantages of ensemble learning to improve the model's performance. We propose a novel strategy: Multi-Label Training, which draws on some advantages of ensemble learning while avoiding the significant increase of computation and storage burden. We use the model proposed by Shon et al \cite{DBLP:conf/interspeech/ShonTG19} as the baseline, and after modifying it based on the proposed strategy idea, we output the final identification results. The results show that the Multi-Label Training further improves the identification rate compared with the baseline.

\section{New method and explanation}
In our proposed method, we split the training data set into \(N\) subgroups, each contains all speakers. However, the speakers in each group correspond to a different set of labels.  

Denote \(x\), \(y\) as the speech and label, and define
\begin{equation}
  D = \left\{ (x_{1},y_{1}),\dots,(x_{n1},y_{n1}) \right\}
  \label{eq1}
\end{equation}
as the training set where \(n1\) is the total number, and define
\begin{equation}
  T = \left\{ (x_{1},y_{1}),\dots, (x_{n2},y_{n2})\right\}
  \label{eq2}
\end{equation}
as the testing set where \(n2\) is the total number. Given a training data set \(D\) and testing data set \(T\) with \(C\) classes, \(D\) is sorted by Speaker ID and \(N\) cannot be bigger than the minimum number of speeches for each speaker. The specific steps are as Table 1. Here the cross-entropy loss function is proposed as 
\begin{equation}
L = -\sum_{c=1}^{N \times C}p_{c}\mathrm{log}(q_{c})
\label{eq3}
\end{equation}
where \(p\)$_c$ refers to the ground truth according to the \(y\) and \(q\)$_c$ refers to the prediction.
\begin{table}[h]
	\caption{Procedures of Multi-Label Training}
	\label{try}
	\centering
	\begin{tabular}{l}
		\toprule
		\textbf{Strategy: Multi-Label Training}             \\
		\midrule
		TRAINING(step 1 to 7):\\
		\textbf{Input:} Training set \(D\) = $\{$(\(x\)$_1$,\(y\)$_1$), $\dots$, (\(x\)$_n$$_1$,\(y\)$_n$$_1$)\}, \(N\), \(C\).\\
        1: \textbf{for} \(i\) = 1, ..., \(n1\) \textbf{do}\\
		2:   \ \ \ \(m\) $\leftarrow$ \(i\) mod \(N\);\\
		3:  \ \ \  \textbf{for} \(m\) = 0, ..., \(N\)-1 \textbf{do}\\
		4:   \ \ \ \ \ \  \(y\)$_i$ $\leftarrow$ \(y\)$_i$ + (\(C\) × \(m\));\\
		5:  \ \ \ \textbf{end for}\\
		6: \textbf{end for}\\
		7: Train the model on updated \(D\);\\
		TESTNING(step 8 to 17):\\
		\textbf{Input:} Testing set \(T\) = $\{$(\(x\)$_1$,\(y\)$_1$), $\dots$, (\(x\)$_n$$_2$,\(y\)$_n$$_2$)\}, \(N\), \(C\).\\
		\textbf{output:} Accuracy.    \\
		8: Set correct=0, total=0;\\
		9: \textbf{for} \(i\) = 1, ..., \(n2\) \textbf{do}\\
		10: \ \ \ Compute the predicted label: \(L\);\\
		11: \ \ \ \textbf{for} \(i\) = 0, ..., \(N\)-1 \textbf{do}\\
		12: \ \ \ \ \ \ \textbf{if} \(L\) == \(y\)$_i$ + (\(C\) × \(i\)) \textbf{do}\\
		13: \ \ \ \ \ \ \ \ \ correct $\leftarrow$ correct + 1;\\
		14: \ \ \ \textbf{end for}\\
		15: \ \ \  total $\leftarrow$ total + 1;\\
		16: \textbf{end for}\\
		17: Accuracy $\leftarrow$ correct ÷ total;\\
		 
		\bottomrule
	\end{tabular}
\end{table}
 
We use the model that Shon et al. \cite{DBLP:conf/interspeech/ShonTG19} proposed as the baseline, where the output layer dimension corresponds to the number of speakers. We modified the baseline based on the proposed strategy and compared it with the baseline. In this strategy, the output layer dimension will change to \(N\) times of the original. Therefore, the classification category becomes \(N\) times of the original, which is equivalent to adding an additional \(N\)-1 classifiers similar to the baseline. So this strategy has some similarities to ensemble learning, such as bagging (bootstrap aggregating), which constructs \(K\) different data subsets by resampling the original data set, trains each model on the corresponding subset, and polls the output in an average way. However, we know that "Model averaging is usually discouraged when benchmarking algorithms for scientific papers, because any machine learning algorithm can benefit substantially from model averaging at the price of added computation and memory \cite{lecun2015deep}". But there is only one model here, and we've only made a few modifications to the FC layer. Although the data set is split into \(N\) subgroups, unlike Bagging, these subsets is not resampled, so they are all one \(N\)th the size of the original data set, but all include all speakers. So obviously there's no significant increase in computation and memory compared with ensemble learning. The way we construct the subsets of data is similar to article \cite{ls/asc/MaoCJGW19}, and the subsets are completely mutually exclusive, increasing the complexity of the ensemble system according to the article \cite{DBLP:journals/eswa/AgarwalC20}. In training, It can be seen as the ensemble of several individual classifiers, each classifier is trained on the corresponding data subsets, the final prediction is made through the maximum vote. Since each group of data is one-\(N\)th of the original data set, the identification rate of each part will certainly fall as a result, because of the insufficient training data. But also because of the compensation of maximum vote and multiple correct labels, the overall identification rate will rise. The reason why the Multi-Label Training may have a better performance, in fact, is that it can reach a better middle performance between the rise and the fall, which is similar to the article\cite{ls/asc/MaoCJGW19}\cite{DBLP:journals/eswa/AgarwalC20}. 
\begin{table*}[h]
	\caption{Performance before and after adding a fully connected layer to the baseline model}
	\centering
	\begin{tabular}{lllll}
		\toprule 
		\multirow{2}{*}{\textbf{Identification Net(Using D)}} & \multicolumn{2}{l}{ \textbf{VoiceID Loss \cite{DBLP:conf/interspeech/ShonTG19}(2 FC Layers) }} & \multicolumn{2}{l}{ \textbf{VoiceID Loss \cite{DBLP:conf/interspeech/ShonTG19}(3 FC Layers) }} \\
		& \textbf{Top1 Accuracy(\%)} & \textbf{Top5 Accuracy(\%)} & \textbf{Top1 Accuracy(\%)} & \textbf{Top5 Accuracy(\%)}\\ 
		\midrule
		\textbf{Clean Speech} & 87.2 & 95.5 & 87.1 & 95.0\\
		\bottomrule
	\end{tabular}
\end{table*}
\section{Model architecture}

As is shown in Figure 1, the model in this paper is modified from the baseline model proposed by Shon et al. \cite{DBLP:conf/interspeech/ShonTG19} and the same as the baseline except for the fully connected layers and the output layer. Compared with the baseline model \cite{DBLP:conf/interspeech/ShonTG19}, the dimension of the output layer was changed to \(N\) times of the original. We cannot avoid modifying the model, so we have to consider how to exclude other interference factors such as the number of model parameters. Here we can draw on the work of Nagrani et al. \cite{nagrani2017voxceleb} and according to the article \cite{DBLP:journals/csl/RohdinSDPMBG20, Rachmatullah2019DeterminingTN}, we believe that the network model is not the deeper the better, and the number of nodes in hidden layers will also have an impact. When conducting controlled experiments to compare the performance of two algorithms, we can make the dimension of the last hidden layer nearly proportional to the output layer dimension, thus excluding interference from structural suitability. In order to verify our understanding, we modified the baseline model proposed by Shon et al. \cite{DBLP:conf/interspeech/ShonTG19} and tested it. The two-layer fully connected layer and the output layer of the baseline model is represented as: 1500--600--1251. In order to exclude the influence of the increasing parameters, we added an additional  fully connected layer(dim=2500) to the baseline model, which became: 1500--2500--600--1251, and the performance is shown in Table 2. We did find that there was little difference between the two. So in this model for speaker identification, we can see that adding or deleting a hidden layer  cannot make a big difference to the model's performance. And we should pay more attention to ensure that the ratio of the dimension between the last hidden layer and the output layer is approximately constant during the experiment.

\subsection{Speech enhancement module}
The speech enhancement model consists of 11 dilated convolution layers. speech enhancement is not the focus of this paper, so we directly use the speech enhancement model proposed by Shon et al. \cite{DBLP:conf/interspeech/ShonTG19}, and its specific structure is shown in Table 3. For the output of the last convolution layer, we use the sigmoid function to generate a ratio mask of the same size and multiply the ratio mask with the original input to achieve speech enhancement, after which the enhanced speech enter the speaker identification model. For the other layers, we use RELUs as non-linear activation. The final cross-entropy loss is calculated by comparing the prediction derived from SOFTMAX with the real identity of the speaker.
\begin{table}[h]
	\caption{Structure of the speech enhancement model}
	\label{tab:word_styles}
	\centering
	\begin{tabular}{lll}
		\toprule
		\textbf{Layer Name} & \textbf{Structure} & \textbf{Dilation}             \\
		\midrule
		Conv1               & 1×7×48             & 1×1                             \\
		Conv2               & 7×1×48             & 1×1                  \\
		Conv3               & 5×5×48             & 1×1             \\
		Conv4               & 5×5×48             & 2×1                 \\
		Conv5               & 5×5×48             & 4×1      \\
		Conv6               & 5×5×48             & 8×1  \\
		Conv7               & 5×5×48             & 1×1\\
		Conv8               & 5×5×48             & 2×2                  \\
		Conv9               & 5×5×48             & 4×4\\
		Conv10              & 5×5×48             & 8×8    \\
		Conv11              & 1×1×1             & 1×1   \\
		
		\bottomrule
	\end{tabular}
\end{table}

\subsection{Speaker identification module}

The speaker identification model uses 1D convolution to consider all bands at once. We use the model proposed by Shon et al. \cite{DBLP:conf/interspeech/ShonTG19} as the baseline of the experiment and we want to modify it for the Multi-Label Training as little as possible without introducing other interfering factors. The speaker identification model of the baseline consists of four 1D convolution layers and two fully connected layers (dimensions: 1500, 600), while the speaker identification model used for Multi-Label Training consists of four one-dimensional convolution layers and one fully connected layer (dimension: 1500). The filter sizes of the four one-dimensional convolution layers are: 5, 7, 1, 1; the strides are: 1, 2, 1, 1; and the number of filters in each layer is shown in Figures 1. There is also a global average pooling layer between the final convolution layer and the fully connected layer. The approach of spectrum extraction here is the same as the speech enhancement model. Finally we extract the speaker embedding from the last fully connected layer. 
\section{Experiments}
\begin{table}[h]
	\caption{Descriptions of the 3 models of the experiments}
	\label{tab:word_styles}
	\centering
	\begin{tabular}{ll}
		\toprule
		\textbf{Model} & \textbf{Description}             \\
		\midrule
		\multirow{2}{*}{VoiceID Loss\cite{DBLP:conf/interspeech/ShonTG19}}    &Using the model proposed by Shon et al.\\ &\cite{DBLP:conf/interspeech/ShonTG19} as the baseline.\\
		\midrule
		\multirow{3}{*}{Proposed(\(N\)=3)}        & Using the strategy proposed in this\\& paper, the original training data set is\\ &split into 3 subgroups. \\
		\midrule
		\multirow{3}{*}{Proposed(\(N\)=2)}       & Using the strategy proposed in this\\ &paper, the original training data set is\\ &split into 2 subgroups.\\

		\bottomrule
	\end{tabular}
\end{table}
On a data set with a limited size, an excessively large \(N\) will cause the data to be too sparse, and ultimately reduce the overall  identification performance. Therefore, in this experiment, we only test the conditions that \(N\) is 2 and 3, in which the ratio of the dimension between the last hidden layer and the output layer is close to the baseline, which is consistent with the conclusion discussed in last section. Thus the model structure in Figure 1 is ideal and we don't need to make any more modification.
\begin{table*}[th]
	\caption{Results of the three models tested with clean speech and various noisy speech}
	\label{tab:example}
	\centering
	\begin{tabular}{llllllll}
		\toprule
		\multicolumn{2}{c}{\textbf{Identification(Using \(D\))}} & 
		\multicolumn{2}{c}{\textbf{VoiceID Loss \cite{DBLP:conf/interspeech/ShonTG19}}} &
		\multicolumn{2}{c}{\textbf{Proposed(\(N\)=3)}} &
		\multicolumn{2}{c}{\textbf{Proposed(\(N\)=2)}}\\
		\midrule
		\multicolumn{2}{c}{\textbf{Enhancement(Using \(D\)$^N$)}}&
		$\times$ & $\surd$& $\times$ & $\surd$ & $\times$ & $\surd$ \\
		\midrule
		\textbf{Type} & \textbf{SNR}   
		&\multicolumn{2}{c}{\textbf{Accuracy(\%)}}
		&\multicolumn{2}{c}{\textbf{Accuracy(\%)}}
		&\multicolumn{2}{c}{\textbf{Accuracy(\%)}}\\  
		\midrule
		clean &--- & 87.2&87.9 &88.4 &89.0 &88.9 & 89.4 \\
		white & 10 & 10.9&71.2 & 6.2 & 73.6 & 5.8& 75.5\\   
		babble & 10 & 43.5&76.8 & 20.6 & 79.9 & 21.3& 80.3\\        
		leopard & 10 & 71.3&82.8 &  62.7 &  84.7 & 63.1& 85.6\\   
		volvo & 10 & 58.1&83.5 &  44.3 &   83.8 & 44.8& 84.7\\    
		factory & 10 & 18.6&70.6 &   7.3 &   72.2 & 8.0& 73.6\\      
		tank & 10 & 52.7&78.7 &   38.1 &  82.4 & 37.6& 84.7\\  
		gun& 10 & 75.0&84.2 &   65.5 &  86.2 & 66.1&  87.0\\  
		
		\bottomrule
	\end{tabular}
	
\end{table*}

\subsection{Data}
In our work, we use the Voxceleb1 \cite{nagrani2017voxceleb} data set, which is text-independent and extracted from YouTube video. It contains 1251 speakers and about 150,000 speech, each with an average length of 7.8 seconds. The total duration of this data set is about 350 hours. Here we directly use the window function with 25ms window size and 10ms shift to take out the 257-dimensional spectrum as input. We do not normalize the input data, just take an exponential multiple of the amplitude spectrum of 0.3. We used 298 frames of fixed length (257 each frame) as input.

To test the robustness of the model, we used the Noise-92 noise data set. Since the Voxceleb1 \cite{nagrani2017voxceleb} data set is taken from YouTube, each speech inevitably contains some noise, but this paper follows the example of paper \cite{DBLP:conf/interspeech/ShonTG19} and take the Voxceleb1 \cite{nagrani2017voxceleb} data set as clean speech(abbreviate training set as \(D\)). We selected seven different types of noise from the Noise-92 data set: White, Babble, Leopard (military vehicle noise), Volvo (vehicle interior noise), Factory, Tank, and Gun (machine gun noise), and mixed them with clean speech at a signal-to-noise ratio of 10 dB to obtain noisy speech(abbreviate noisy training set as \(D\)$^N$) for training and testing model in noisy conditions.

\subsection{Speaker identification}

When speaker identification was performed, both the training set and the test set contain all speakers (1251). And the baseline model and the proposed models in this paper should have the same size of training set and test set. In this paper, we just test the performance of the Multi-Label Training when \(N\)=2 and 3. For VoiceID Loss \cite{DBLP:conf/interspeech/ShonTG19}, we divided the original data set into training and test sets in a 3:1 ratio. For Proposed (\(N\)=3) and Proposed (\(N\)=2), we divided the original data in the same way first, and then we followed the procedures as in Table 1 and test the performance of the Multi-Label Training.

When mixing noisy data with clean speech, the noisy data is first split into two parts, one for mixing with the training set data and one for mixing with the test set data, so that the same noise data is not used by both the training set and the test set.

\subsection{Result and discussion}
The experimental results are the average values obtained after two repeated experiments and are shown in Table 5 and Figure 2. Compared with baseline, the proposed strategy improves the performance at both \(N\) for 2 and 3, although it is better at \(N\) for 2. So it seems that it is more appropriate to splitting the data set into two subgroups on Voxceleb1. If testing on a more dense data set, perhaps in the best case, \(N\) can be a larger number. 
\begin{figure}[h]
	\centering
	\includegraphics[width=65mm]{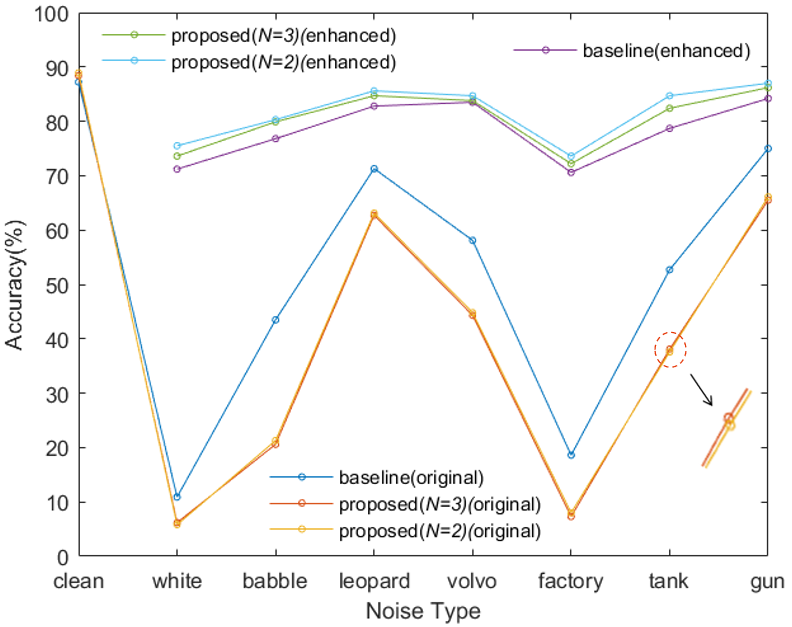}
	\caption{Line graph of the results}
	\label{result1}
\end{figure}
It can also be found that the identification rate deteriorates significantly when using the noisy speech to directly test identification model without enhancement, and the performance of the model trained by the proposed strategy drop even more. The reason should be that after splitting into two or three subgroups, each group is reduced to one-half or one-third of the original training set, the corresponding part is trained insufficiently as a result. When inputting the noisy speech directly, the performance of each part drops sharply. Although the final result is compensated by maximum vote and multiple correct labels, the overall performance still decreases. However, we found that after the speech enhancement, the performance of the two proposed models are still significantly improved compared with the baseline, indicating that our proposed strategy can still improve the performance of the model in both clean and noisy conditions. Comparing the results before and after enhancement, it can be seen that the speech enhancement model is very helpful for improving the system's robustness. And by comparing the performance between different types of noise, we can find that the noise of wide frequency domain interferes the model more strongly. For example, white noise has a greater impact on the model than babble. Maybe because the goal of speech enhancement here is to improve the identification rate, rather than sound more clearly, so this also shows that this model is actually very different from the human auditory senses. 

\section{Conclusions}
In this paper, we propose a novel strategy: Multi-Label Training, which divides the training data set into several mutually exclusive subgroups and sets the corresponding identity labels. Then it modifies the dimensions of the output layer accordingly to achieve the effect of a pseudo ensemble learning with very low additional complexity added. Compared with the use of various superior model structures in speaker identification, the Multi-Label Training chooses another way. It can be easily applied to almost all text-independent speaker identification models like ensemble learning, thereby further improving the performance. This paper just uses the model proposed by Shon et al \cite{DBLP:conf/interspeech/ShonTG19} as the baseline and the results show that the Multi-Label Training can further improve the performance of the model in both clean and noisy conditions.

\bibliographystyle{IEEEtran}
\bibliography{mybib}

\end{document}